\documentclass{elsart}
\usepackage{graphicx}

\def\url#1{{\ttfamily\def\/{/\discretionary{}{}{}}#1}}

\def\ltsima{$\; \buildrel < \over \sim \;$}
\def\simlt{\lower.5ex\hbox{\ltsima}}
\def\gtsima{$\; \buildrel > \over \sim \;$}
\def\simgt{\lower.5ex\hbox{\gtsima}}

\begin{document}

\begin{frontmatter} 
\title{Mapping the circumnuclear matter of NGC~1068 in X-rays}
\author{M. Guainazzi$^1$}
\address{Astrophysics Division, Space Science Department of ESA, ESTEC/SA, Postbus 299, NL-2200 AG, The Netherlands}
\author{S. Molendi, P. Vignati}
\address{Istituto di Fisica Cosmica ``G. Occhialini'', C.N.R., Via Bassini 15, I-20133 Milano, Italy}
\author{G. Matt}
\address{Dipartimento di Fisica, Universit\`a degli Studi ``Roma Tre",
Via della Vasca Navale 84, I--00146 Roma, Italy}
\author{K. Iwasawa}
\address{Institute of Astronomy, University of Cambridge, Madingley Road, Cambridge CB3 0HA, United Kingdom} 
\thanks[email]{Current address: XMM Science Operation Center, VILSPA, ESA, Apartado 50727, E-38080 Madrid, Spain. E-mail: mguainaz@xmm.vilspa.esa.es }
\begin{abstract} 
An-energy dependent variation of the X-ray emission from the
archetypical ``Compton-thick'' Seyfert~2 galaxy NGC~1068 has been observed
between two BeppoSAX observations, performed one year apart. This variation
(at the 2.6$\sigma$ level) is highest
in the 3--10~keV band and declines towards lower and higher energies. The
most straightforward explanation is a variation of the
primary nuclear continuum, which is scattered along our line of sight
by an electron plasma.
If this is indeed the case, this evidence allows us to obtain the
first {\it direct} estimate of
the location of the scattering medium, at $\sim$1~pc from the nucleus.
\end{abstract} 
 

\end{frontmatter} 

\section{Introduction}

Compton--thick Seyfert 2 galaxies,
i.e. sources in which the nucleus is obscured by matter with column
densities exceeding $\sim \sigma_{th}^{-1} \simeq 
10^{24}$~cm$^{-2}$, can be observed, in X--rays, only
in scattered light. In the best known case, NGC~1068, at least two
scattering components are known to be
present (e.g. Iwasawa, Fabian \& Matt 1997;
Matt et al. 1997). One originates from almost neutral material,
possibly the inner
surface of the molecular torus, envisaged by the Seyfert unification scenarios,
Antonucci \& Miller 1985; Antonucci 1993. It is
recognizable by the characteristic shape of the continuum, peaking 
between 20 and 30~keV, which is
determined by the competition between photoabsorption and Compton scattering.
A prominent iron line at 6.4 keV comes along with it (George \& Fabian 1991; 
Matt, Perola \& Piro 1991). 
The other originates from ionized matter, and is a fainter replica of the
nuclear radiation, at least as long as self-absorption effects are negligible.
Lines from ionized elements, most
commonly the He-- and H--like stages, are superposed to the latter. 

Compton--thick Seyfert 2s are therefore ideally suited for studying 
the circumnuclear matter in AGN. From X--ray spectroscopy, much can 
be learned on the physical and chemical properties of the reflecting
matter (e.g. Netzer 1996; Matt, Brandt \& Fabian 1996, Netzer \& Turner 1998).
Even if a full exploitation of the line diagnostic capabilities
should await high spectral resolution instruments, first
results have already been obtained with moderate resolution detectors
(Netzer \& Turner 1998; Guainazzi et al. 1999, G99 hereinafter).
To map the location of the 
reflectors, instead, variability measurements are required. 
While an accurate mapping requires a good sampling performed with
a high sensitivity detector (and even in this case a real reverberation
technique cannot be applied due to the fact that the nuclear 
radiation is not directly observable),
basic pieces of information can be derived by the crude
comparison of observations  
made at different epochs, preferentially with the same
instrument to avoid cross--calibration problems.

The Italian-Dutch X--ray satellite
BeppoSAX  (0.1--200 keV, Boella et al. 1997a) observed
NGC~1068 twice, about one year apart. We therefore
searched for variability between the two observations, and found an energy
dependent one. In Sect.~2 a detailed description of the data reduction is
presented, and possible systematic effects are discussed in Sect.~3. As the variability
survived all our checks, in Sect.~4 we discuss the results and show that
a possible explanation of the broadband X-ray variability
is in term of a flattening of the nuclear
component, mirrored by the ionized reflector. In Sect.~5 we derive some geometrical
and physical properties of the warm scatterer (and of the other spectral components)
implied by our results.

\section{Observation and data reduction}

NGC~1068 was observed by BeppoSAX on December 1996 and January 1998
(see Tab.~\ref{tab1}).
\begin{table}
\caption{NGC~1068 BeppoSAX observation log}
\vspace{0.25 cm}
\begin{tabular}{lccccc} \hline \hline
ID & Start Time & End Time  & ${\rm T^{LECS}_{exp}}$ & ${\rm T^{MECS}_{exp}}$ &
${\rm T^{PDS}_{exp}}$ \\
 & (UTC) & (UTC) & (ks) & (ks) & (ks) \\ \hline
December 1996 & 30/12/1996 08:47:25 & 03/01/1997 05:27:50 & 61497 & 100150 & 62493 \\
January 1998& 11/01/1998 09:52:36 & 12/01/1998  08:07:50 & 15408 & 37331 & 17657 \\ \hline \hline
\end{tabular}
\label{tab1}
\end{table}
Of the four co-aligned instruments, comprising the BeppoSAX scientific payload,
we will deal in this {\it paper}
with data from the Low Energy Concentrator Spectrometer (LECS, 0.1--10~keV,
Parmar et al. 1997), the Medium Energy Concentrator Spectrometer (MECS, 1.8--10.5~keV,
Boella et al 1997b) and the Phoswitch Detector System 
(PDS, 13--200~keV; Frontera et al. 1997).
LECS and MECS are imaging gas proportional counters, which use identical concentrator systems, with moderate
energy resolution ($\simeq$8\% at 6~keV).
The MECS effective area in the two units configuration discussed in this
{\it paper} is about twice as the LECS one in the overlapping energy band.
The PDS consists of four independent
crystal units arranged in pairs, each mounted on a rocking collimator to achieve
a continuous monitoring of the background with a duty-cycle of 96~s.

Data reduction followed standard procedures, as described in G99. In particular,
Good Time Intervals
(GTI) for scientific products accumulation were defined according to the following
selection criteria: {\it a)} the angle between the pointing directions and the
Earth's limb
was higher than 4$^{\circ}$; {\it b)} passages through the South Atlantic
Geomagnetic Anomaly were excluded;
{\it c)} the angle between the pointing direction and the direction of the Sun was greater
than
60$^{\circ}$. Count rates 
were extracted from circular regions of radius 8' and 4' around the NGC~1068 centroid
in the LECS and MECS, respectively. No other point
source is detected in these areas at a level higher than $6 \times 10^{-14}$
and $1.6 \times 10^{-13}$~erg~cm$^{-2}$~s$^{-1}$ (3 $\sigma$ local background
fluctuations)
in the 0.1--2~keV and 2--10~keV energy bands, respectively. For comparison,
G99 report average fluxes of 1.1 and $0.5 \times 10^{-11}$~erg~cm$^{-2}$~s$^{-1}$
for the BeppoSAX observations of NGC~1068 in the same bands. PDS data were 
additionally screened by removing
five minutes intervals after any SAGA passage, to allow gain recovery to nominal
values after instrumental switch-offs.

We have performed an analysis of several BeppoSAX Crab observations
to estimate if aging effects in the detectors could produce a secular variation of the observed
count rate. No effect of this kind is observed, the upper limit being 0.3\% and 0.7\% for the
LECS and MECS, respectively. The count rates
obtained from the two NGC~1068 observations may be
hence directly compared.

\section{LECS data reduction issues}

One of the most critical issues in estimating the variability of a relatively
faint source as
NGC~1068 is the background subtraction. In Tab.~\ref{tab2} we compare the count
rates in three energy bands (0.1--1~keV; 1--3~keV; 3--10~keV) when
the three background subtraction techniques
described in Parmar et al. (1999) are employed.
They are:  {\it a)} background extracted from blank sky deep
exposures,
accumulated by the BeppoSAX Science Data Center (SDC) in the first three years 
of operative life of the mission (BLS);
{\it b)} background extracted from two semi-annuli in the LECS field of view, suitably
renormalized
to the expected counts under the source extraction region (FOV); 
{\it c)} cosmic background estimated from the 
7-bands ROSAT/PSPC All Sky Survey source-removed count rates (RM), and added
to the instrumental background.
\begin{table}
\caption{LECS background subtracted count rates (in units of $10^{-2}$~s$^{-1}$) when the
three different techniques of background subtraction after Parmar et al. (1999) are adopted
(details in text)}
\vspace{0.25cm}
\begin{center}
\begin{tabular}{lccc} \hline \hline
& 0.1--1~keV & 1--3~keV & 3--10~keV \\ \hline
\multicolumn{4}{l}{December 1996} \\
BLS & $6.29 \pm 0.10$ & $4.07 \pm 0.08$ & $1.51 \pm 0.07$ \\
FOV & $6.58 \pm 0.11$ & $4.02 \pm 0.09$ & $1.37 \pm 0.07$ \\
RM & $6.28 \pm 0.11$ & $4.00 \pm 0.09$ & $1.55 \pm 0.08$ \\
\multicolumn{4}{l}{January 1998} \\
BLS & $5.82 \pm 0.20$ & $4.01 \pm 0.17$ & $1.60 \pm 0.13$ \\
FOV & $6.10 \pm 0.20$ & $3.96 \pm 0.18$ & $1.45 \pm 0.13$ \\
RM & $5.81 \pm 0.20$ & $3.94 \pm 0.18$ & $1.64 \pm 0.13$ \\ \hline \hline
\end{tabular}
\end{center}
\label{tab2}
\end{table}
The results differ by about 5\% , 2\% and 12\% in the three bands,
respectively. 
This differences might be due to local granularity in the
contribution
of the cosmic background, which is globally mapped on scales not lower than
2$^{\circ}$
(Snowden et al. 1995). On the other hand, the instrumental LECS background is known to
exhibit
secular variations with a dynamical range of about a factor two (T.Oosterbroek, private communication).
It is therefore impossible
to tell which of the three techniques is giving the "true" result
(see also the discussion in Parmar et al. 1999). Although the count rates uncertainties
are dominated by systematic effects, the ratio between the count rates observed in
the two BeppoSAX observations is independent of the background subtraction
method used. Assuming the BLS one (the same adopted normally in spectral fitting),
the ratio of the January 1998 versus the December 1996 count rates are: $0.93 \pm 0.04$, and
$0.99 \pm 0.05$ and $1.06 \pm 0.11$ in the 0.1--1~keV, 1--3~keV and
3--10~keV energy bands,
respectively.

Count rate variations by a few percent could be ascribed to the a different positions of
the source in the LECS field of view. It is in facts crossed
by a mesh of fine grids, which cause a $\sim$5\% modulation
of the count rates on scales $\sim$10'', and by the strongback support rib, which
causes a $\simeq$30\% absorption at regular distances of 2' (along both the X and Y
axis of the focal plane). The former effect has a much lower spatial scale
than the PSF width ($\simeq$4' at 0.28~keV; $\simeq$0.9' at 6~keV), and is therefore
averaged out when wide enough extraction radii are taken.
The latter effect is unlikely to introduce a significant variability.
The best-fit centroid of the source profiles differ by $\simeq$40" between the two BeppoSAX observations.
On this scale, we estimate a count rate difference $\simlt 1.5\%$, well within the statistical uncertainties of our measures.

Response matrices appropriated for the source position in the two
observations were extracted and used in the joint LECS/MECS spectral fitting. The
results of the spectral fittings, described
in detail in Appendix below, confirm the above outcomes.

\section{MECS data reduction issues}

In Tab.~\ref{tab5} we report the MECS background subtracted 
\begin{table}
\begin{center}
\caption[]{MECS background subtracted count rates (in units of $10^{-2}$~s$^{-1}$)}
\vspace{0.25cm}
\begin{tabular}{ccc} \hline \hline
1.5--3.0~keV & 3.0--5.0~keV  & 5.0--10.0~keV\\ \hline
\multicolumn{3}{l}{December 1996} \\
$2.40 \pm 0.05$ & $1.44 \pm 0.04$ & $2.36 \pm 0.05$ \\
\multicolumn{3}{l}{January 1998} \\
$2.39 \pm 0.08$ & $1.63 \pm 0.07$ & $2.57 \pm 0.09$ \\ \hline \hline
\end{tabular}
\end{center}
\label{tab5}
\end{table}
count rates for the two observations of NGC 1068 in the
1.5--3~keV, 3--5~keV and 5--10~keV energy bands.
The softest ratio is consistent with unity,
while for ${\rm E > 3}$~keV the ratio 
is larger than 1 at the 2.6$\sigma$ level.

The change in the position of the 
source within the MECS detector from one pointing to the other
(constrained by the better MECS statistics to be
$\simlt$30'') implied a difference in the vignetting
$\simlt$ 1\%.
The gain is stable between the two observations. The difference
between the best-fit centroids of the prominent iron line in the
channel space between the two observations
is $0.6 \pm 0.9$~PI (1 channel corresponds to
$\sim$50~eV).
We note that the MECS does not suffer of any support grid
mesh obscuration problem. A support ring
is present at about 10' from the center of the field
of view and is almost concentric with it. No change of the
MECS effective area on arc minute scales is therefore expected,
other than that ascribed to the instrumental vignetting.
Background subtraction has been performed by using 
blank field observations. 
To verify that the background during the two
NGC 1068 observations is stable and consistent
with that obtained from the blank fields, we 
have accumulated the count rates from the outer region 
of the detectors, where the contribution from the 
source is negligible. We find that the
background count rates vary by less than 
$\sim 6$\% between the two observations.
Considering that the background count rate 
contributes $\sim 8$\% to the source plus
background count rate,
we estimate that any spurious source count rate variation 
induced by inaccurate background subtraction will 
be smaller than $\sim$0.5\%. A systematic error of 1.5\% will
be hereinafter conservatively added
to the MECS count rates to account for possible vignetting estimate
and background subtraction inaccuracies.

\section{Comparison between the broadband spectra}

\begin{figure}
\caption{Count rate ratio between the January 1998 and December 1996 observations
as a function of energy. {\it Empty squares}: LECS; {\it filled circles}: MECS;
{\it cross}: PDS. The {\it dotted} lines indicate the MECS measures in the
3--5~keV and 5--10~keV separately.
The {\it solid} and {\it dashed} line in the PDS data point
represent the statistical and statistical plus systematic uncertainties after
Guainazzi \& Matteuzzi (1997). The 3--10~keV LECS measure ($1.06 \pm 0.11$)
is consistent with the MECS one and not shown for sake of clarity}
\begin{center}
\includegraphics*[width=9.0cm,height=9.0cm]{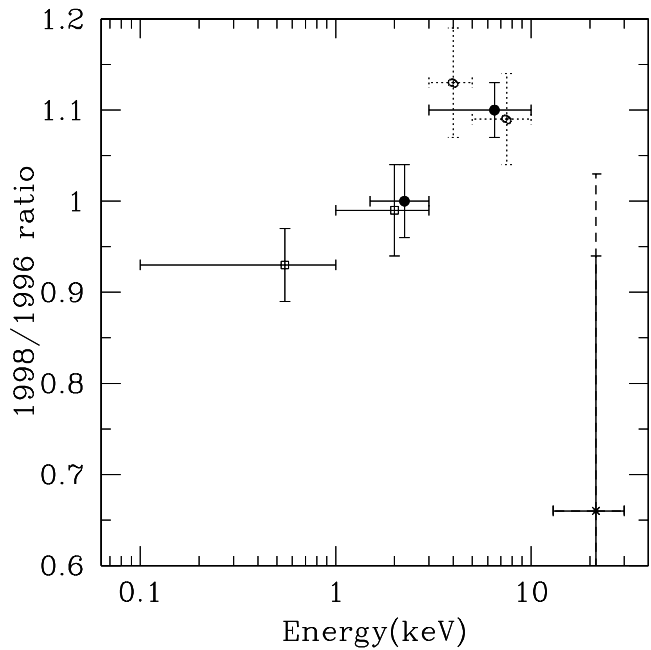}
\end{center}
\label{fig1}
\end{figure}

In Fig.~\ref{fig1} we show the ratio
between the January 1998 and the December 1996 spectra.
This plot includes also the ratio of the PDS count rates in the 13--30~keV energy
band. As noted before, the count rate ratio is
$\sim$0.9 below 1~keV, becomes 
consistent with unity in the 1--3~keV band, reaches
a maximum ($1.098 \pm 0.038$) between 3 and 10~keV
and then decreases again, returning consistent
with 1 above 15~keV.

The maximum of the variability occurs in the energy
range where the contribution of the warm scattered component is the highest
(G99).
The amount of the observed variability
is admittedly small, and the spectral complexity hampers a clear
identification of  which component(s) is responsible for it. However,
a comparison of Fig.~\ref{fig1} with G99 suggests that
the observed variability might be due to
a flattening of the warm scattered component around
a pivot point located somewhere in the 1--4~keV band. More complex scenarios cannot
be ruled out, and the interplay of spectral steepness and normalizations
can further contribute to confuse the picture. Variability measurements of
the lines associated with one or both the reflection components should help 
enlightening on the real behavior, but for this purpose much higher
spectral resolution detectors are required.

In Appendix, we show the comparison between the fitting results for the
December 1996 and January 1998 spectra.

\section{Discussion}

We report on the observation of an energy dependent flux variability 
in NGC~1068 on timescales of
about one year. The flux ratio of the 1998 over the
1996 observations is higher than 1 between 3 and 10~keV,
and decreases towards 
lower and higher energies.
Although the result is significant at the 2.6$\sigma$ level only, its
possible implications are rather far-reaching, and we cannot refrain
to discuss them in this Section. The next generation
of high throughput, high resolution X-ray detectors will allow a
robust testing of the above results and of the forthcoming hypothesis.

General arguments on the spectral 
deconvolution suggest that the observed variation
is best explained in terms of a change
of the warm reflection component, echoing the same variation in the nuclear 
spectrum. The statistics is not good enough, however, to tell whether it is the
spectral index or the normalization of the warm-scattered component (or both),
that is responsible for the observed variability (see Appendix). Similarly,
our data cannot test independently the hypothesis
that self-absorbing effects in the warm-scattered component are negligible.

Apart from the details, it is rather plausible that the measured variation
is mainly due to the warm reflection component. This may permit
to derive information on the size and location of the reflecting matter.
Light crossing arguments suggest that the size of the region cannot exceed
one light--year, i.e one--third of a parsec. As the light--curve 
of the nuclear emission is unknown, we cannot determine with certainty
the distance of the reflector to the nucleus. However, an
order--of--magnitude estimate may be derived by recalling that the
2--10 keV luminosity of the warm reflection component is about
1.5$\times10^{41}$ erg s$^{-1}$, and that its Thomson optical depth, $\tau_T$,
 is estimated to be a few$\times10^{-3}$ (G99). The relation 
between nuclear and reflected luminosity may be written as:

\begin{equation}
L_{ref} = \tau_T f L_{nuc} 
\end{equation}

\noindent
where $f$ is a geometrical factor basically equal to the solid angle 
subtended by the reflecting matter to the nucleus, in units of 4$\pi$. 
Therefore, if the nuclear luminosity is not much larger than 10$^{44}$
erg s$^{-1}$ (Iwasawa et al. 1997), $f$ cannot be much lower than 1.
If this is the case, 
the distance of the reflecting region from the nucleus cannot be much
larger than its size, and should therefore be of the order of a parsec or
less. 
This crude estimate puts the warm scattering medium in an intermediate location
between the Broad (Peterson et al. 1991; Stirpe et al. 1994; Dietrich et al. 1998)
and the Narrow Line Regions (Axon et al. 1997). This supports the idea that
this medium coincides with that responsible for the almost energy-independent
scattering of the nuclear continuum, observed as the appearance of broad
lines in the polarized light optical spectra of NGC~1068 (Antonucci \& Miller
1985). Because in this source the torus is probably observed almost edge--on
(Greenhill et al. 1997; Matt et al. 1997), the warm reflector, in order
to be visible,  must be placed outside the torus itself. This implies a
rather compact torus, with inner side of order of 1 pc or less. This is
in good agreement with the radio continuum measurements, which suggest
the presence of a sub-parsec edge-on disk around the NGC~1068 nucleus
(Gallimore et al. 1998).

Netzer \& Turner (1998) and G99 suggested that the bulk of the ionized
emission lines observed in the soft X-ray spectrum of NGC~1068 originates in
the same medium. The fact that transitions from He-like stages of
elements from oxygen to iron are present, suggests that the ionization
and/or geometrical structure of the scattering medium must be complex. This
would be confirmed by the fact that, while most of the lines do not change
significantly between the two observations, 
the K$_{\alpha}$ Ne{\sc xi} varies by a factor two (see Appendix). 
Given the instrumental and statistical limits of our measurements,
we refrain from pursuing this point any further. This 
results, however, indicates how profitable the study of the line 
dynamics could be,
when the new generation of high resolution/high throughput detectors will be
operative. New constraints on the physical conditions of the matter in the
circumnuclear AGN environment can be set if monitoring with a sufficiently 
fine sampling will be available.

Comparatively little can be said on the variability of the other spectral
components (see Appendix).
The Compton-reflected continuum is detected only in the longer
December 1996 observation. However, the January 1998
upper limit on its 5--200~keV flux
is not inconsistent with the December 1996 one. The
thermal component parameters are remarkably constant. Again, this strengthens
the idea that this component is connected to the strong starburst ring of
approximately 1~kpc size [{\it i.e.}: about 15'' at the distance of 14.4~Mpc
(Tully 1988)], which protrudes a bar towards the nucleus (Scoville 1988).
Actually, half of the emission at 0.8 keV comes from an extended region
of $\simeq$13~kpc scale (Wilson et al. 1992). It is therefore straightforward
to think that the BeppoSAX soft X-ray spectrum is dominated by the same
component.

\section*{APPENDIX -- Time-resolved spectroscopy: an exercise}

We have tried to separately fit the best model of G99 to the December 1996 and
January 1998 LECS/MECS/PDS spectra. In principle this procedure
could allow us to determine exactly which spectral component(s) is responsible
for the observed variability. In practise this is a very hard task, although BeppoSAX at least has
the advantage, over previous missions, of the broad energy coverage which 
allows the best modeling of the continuum so far. We report the relevant results
hereinafter mainly for sake of completeness.
We extracted the MECS spectra from a region of 6' to achieve a better
signal-to-noise ratio. The continuum model is characterized by the spectral
index and normalization of the warm scattered (${\rm \Gamma_{ws}}$, ${\rm N_{ws}}$)
and Compton-reflected (${\rm \Gamma_{cr}}$, ${\rm N_{cr}}$) primary continuum
components, and by the temperature, abundance and normalization of the thermal
component (${\rm kT}$, ${\rm Z}$, ${\rm N_{th}}$; the {\tt mekal} model
in {\sc Xspec} has been used). The Galactic contribution to the photoelectric
absorption has been held fixed to the value $2.9 \times 10^{20}$~cm$^{-2}$,
following Murphy et al. (1996), which is consistent with the value obtained in both
the December 1996 and January 1998 fits
if left as a free parameter.
The results are summarized in
Tab.~\ref{tab3} and Tab.~\ref{tab4} for the continuum and emission lines, respectively.
\begin{table}
\caption{Best-fit results with the baseline G99 model for the December 1996 and January
1998 BeppoSAX NGC~1068 observations}
\vspace{0.25cm}
\begin{center}
\begin{tabular}{lccc} \hline \hline
Parameter &1996 December & \multicolumn{2}{c}{January 1998} \\ \hline
${\rm \Gamma_{ws}}$  & $2.09 \pm 0.16$ & $1.26 \pm^{0.28}_{0.13}$ & $1.8 \pm 0.3$ \\
${\rm N_{ws}}$$^a$ ($10^{-4}$~photons~s$^{-1}$~cm$^{-2}$) & $8.3 \pm^{1.0}_{0.7}$ & $5.0 \pm_{1.6}^{1.0}$ & $7 \pm^4_3$ \\
${\rm \Gamma_{cr}}$ & ${\rm \equiv \Gamma_{ws}}$ & 2.1$^b$/1.25$^b$ & 2.09$^b$ \\
${\rm F^{5-200 \ keV}_{cr}}$ ($10^{-11}$~erg~s$^{-1}$~cm$^{-2}$) & $3.2 \pm^{1.7}_{1.5}$ &  $< 1.7$/$<2.9$ & 3.2$^b$ \\
${\rm kT}$ (eV) & $390 \pm 40$ & $390 \pm^{50}_{60}$ & $380\pm^{60}_{80}$ \\
${\rm Z}$ (\%)  & $1.6 \pm_{0.6}^{0.8}$ & $< 1.3$ & $< 1.9$ \\
${\rm N_{th}}$$^c$ ($10^{-2}$~s$^{-5}$) & $5.4 \pm 0.8$ & $5.7 \pm^{1.4}_{1.2}$ & $5.3 \pm^{1.5}_{1.3}$ \\ 
${\rm \chi^2/}$~dof & 365.6/333 & 150.0/156 & 156.2/156 \\ \hline \hline
\end{tabular}
\end{center}

\noindent
$^a$normalization at 1~keV of the warm scattered component

\noindent
$^b$fixed

\noindent
$^c$normalization of the thermal component
\label{tab3}
\end{table}
\begin{table}
\caption{Emission line best-fit parameters. The continuum models are those in
columns~2 and 3 (for ${\rm \Gamma_{cr} = 1.25}$)
in Tab.~4 for the December 1996 and January 1998 observations, respectively.
Equivalent Width (EW)
are calculated against the warm scattered continuum only, except the Fe{\sc i}, which is given
against the Compton-reflected continuum only}
\vspace{0.25cm}
\begin{center}
\begin{tabular}{lccc} \hline \hline
Line & E & I & EW \\
& (keV) & (photons~cm$^{-2}$~s$^{-1}$) & (eV) \\ \hline
\multicolumn{4}{l}{December 1996} \\
Ne{\sc ix} & $0.95 \pm^{0.02}_{0.03}$ & $(4.2 \pm^{1.1}_{1.2}) \times 10^{-4}$ & $570 \pm ^{150}_{160}$  \\
Si{\sc xiii} & $1.88 \pm^{0.03}_{0.02}$ & $(8.6 \pm 1.3) \times 10^{-5}$ & $460 \pm 70$  \\
S{\sc xv} & $2.46 \pm 0.04$ & $(3.1 \pm^{0.6}_{0.7}) \times 10^{-5}$ & $280 \pm^{50}_{60}$ \\
Fe{\sc i} & 6.4$^a$ & $(5.3 \pm^{0.7}_{0.8}) \times 10^{-5}$ & $1400 \pm^{700}_{800}$ \\
Fe{\sc xxv} & $6.7$$^a$ & $(5.3 \pm^{1.1}_{1.0}) \times 10^{-5}$ & $3500 \pm 800$ \\
Fe{\sc xxvi} & $6.96$$^a$ & $(2.1 \pm^{0.9}_{0.7}) \times 10^{-5}$ & $1500 \pm ^{600}_{500}$ \\ \hline
\multicolumn{4}{l}{January 1998} \\
Ne{\sc ix} & $0.91 \pm ^{0.04}_{0.02}$ & $(8 \pm^3_2) \times 10^{-4}$ & $1500 \pm ^{600}_{400}$ \\
Si{\sc xiii} & $1.84 \pm 0.05$ & $(8 \pm ^3_2) \times 10^{-5}$ & $380 \pm^{140}_{90}$ \\
S{\sc xv} & $2.45 \pm^{0.06}_{0.05}$ & $(5.0 \pm^{1.4}_{1.3}) \times 10^{-5}$ & $330 \pm 90$ \\
Fe{\sc i} & 6.4$^a$ & $(6.0 \pm^{1.4}_{1.5}) \times 10^{-5}$ & $> 2200$ \\
Fe{\sc xxv} & $6.7$$^a$ & $(5 \pm^4_2) \times 10^{-5}$ & $1400 \pm^{700}_{800}$ \\
Fe{\sc xxvi} & $6.96$$^a$ & $(3.4\pm^{1.5}_{1.6}) \times 10^{-5}$ & $1000 \pm^{400}_{600}$  \\ \hline \hline
\end{tabular}
\end{center}

\noindent
$^a$fixed

\label{tab4}
\end{table}
Uncertainties are shown at 1 $\sigma$ level for two interesting parameters ($\Delta \chi^2 =
2.30$) for all the spectral components, except for the thermal optically thin plasma, for
which the values corresponding to three interesting parameters are reported ($\Delta \chi^2
= 3.53$).
In the December 1996 observation,
${\rm \Gamma_{ws}}$ and ${\rm \Gamma_{cr}}$ turned out to be mutually consistent, and therefore
we report in Tab.~\ref{tab4}
the results when these parameters are tied together. The only significant
difference between the two fits is that the warm scattered component is much flatter
and brighter in January 1998,
the 2--10~keV flux increasing from $(1.84 \pm^{0.22}_{0.18})$ to
$(3.4 \pm^{0.7}_{1.1}) \times 10^{-12}$~erg~cm$^{-2}$~s$^{-1}$.
This suggest that it has pivoted around about 1~keV between
the two observations.

The Compton reflected component is actually detected only in the
December 1996 observation, which has by far the higher signal-to-noise ratio,
even if the upper limit of the 5--200~keV flux
in the second observation is consistent with the value
in the first one (Table~5).
The clear presence of the 6.4 keV iron line
also in the second observation (see below) strongly argues against the
disappearing of the cold reflection.
If the CR component is forced
to be the same, in shape and flux, as in the first observation, it is actually impossible to
tell whether it was
the spectral index or the normalization of the WS component that changed
(see column 4 of tab.~\ref{tab4}). 

In Tab.~\ref{tab3}, we show the best-fit results for the emission lines, when the
continuum models of columns~2 and 3 of Tab.~\ref{tab3} are employed.
Most of them do not exhibit a variation 
of either the
best-fit centroid energy or the intensity/Equivalent Width (EW). The only 
remarkable
exception if the Ne{\sc xi} one, whose flux doubles when the warm 
scattered continuum flattens,
yielding an increase of the EW by a comparable factor. On the other hand, 
the ionized iron lines do not
show any significant variation of the intensity, while the underlying 
warm scattered continuum
increased by a factor about two.  This would imply a decrease of their EW 
by the same factor, rather surprising if the continuum and the lines
are produced in one and the same medium. No problems of this kind are
present with the fit of column~4 of Tab.~\ref{tab3}, however, which is therefore again
to be preferred on plausibility grounds.

\section*{ACKNOWLEDGMENTS}

BeppoSAX is a joint Italian-Dutch program.
MG acknowledges the receipt of an ESA Research Fellowship. An anonymous referee
is gratefully acknowledged for a careful scrutiny of the manuscript and several
comments, which greatly improved the quality of the paper.

\end{document}